\documentclass[aps,prl,preprint,superscriptaddress,showpacs,amsmath]{revtex4-1}

\usepackage{graphicx}

\begin{document}

\title{Impact of surface phenomena on direct bulk flexoelectric effect in finite samples}

\author{A.S.~Yurkov}
\affiliation{644076, Omsk, Russia, e-mail:fitec@mail.ru}

\author{A.K.~Tagantsev}
\affiliation{Ceramics Laboratory, Swiss Federal Institute of Technology (EPFL), CH-1015 Lausanne, Switzerland}
\affiliation{Ioffe Phys.-Tech. Institute, 26 Politekhnicheskaya, 194021, St.-Petersburg, Russia}

\date{\today}
\begin{abstract}
In the framework of a continuum theory, it is shown that the direct flexoelectric response of a finite sample essentially depends on the surface polarization energy, even in  the thermodynamic limit where the body size tends to infinity.
It is found that a modification of  the surface  energy can lead to a change of the polarization response by  a factor of two.
The origin of the effect is an electric field produced by surface dipoles induced by the strain gradient.
The unexpected sensitivity of the polarization response to the surface energy in the thermodynamic limit is conditioned by the fact that
the moments of the surface dipoles may scale as the body size.
\end{abstract}
\maketitle 

The impact of the sample surface on  the macroscopic properties of ferroics is an issue that has been attracting the attention of workers during the past 50 years. 
Starting from  the seminal papers for magnetic  \cite{kaganov1972} and non-magnetic \cite{kretschmer1979prb} ferroics, which were based on Landau theory, this issue  latter became a matter of extensive first principles studies (see e.g. \cite{Junquera2003}, \cite{Stengel2009}). All obtained  results  suggest a pronounced size effect, i.e.  an increasing of sample surface  impact  on its macroscopic properties  while decreasing the sample size. Quantitatively, strong effects were predicted for samples having small dimension not exceeding tens of nanometers.

A good example of such phenomena is the out-of-plane dielectric response of ferroelectric thin films. Here, depending on  the film thickness, the effective dielectric constant  \cite{note1} acquires a relative correction of the order of ${\cal L}/h$, where $h$ is the film thickness and ${\cal L}$ is the microscopic scale, which typically does not exceed a few tens of nanometers (see review paper \cite{Tagantsev2006} and references therein).

An essential feature of such size effect is that its strength is, as expected, sensitive to the properties of the surface of the sample, specifically to the value  of the polarization dependent contribution to  the surface energy. 
For example, if such a contribution vanishes (the so-called free boundary conditions) then the aforementioned correction vanishes as well.  In the thermodynamic limit $h\rightarrow\infty$ this correction vanishes also.

The trends in the size effect outline above are of interest not only from the point of view of fundamental science, they also provide guidelines for down-scaling of microelectronic and micro-mechanical devices. In this context, the verification of these trends for  electromechanical effects in solids seems to be a problem deserving a theoretical treatment.
In this letter we present a simple modelling of the impact of the surface polarization energy on the average polarization response of a plate to its bending, which is associated with the bulk flexoelectric effect. We show that, in contrast to the dielectric response, a modification of  the surface polarization energy can readily lead to a 100 $\%$ modification  of the flexoelectric response of a finite sample, which holds in the thermodynamic limit $h\rightarrow\infty$ .
Particularly, we  demonstrate that, at $h\rightarrow\infty$, a modification of  the surface polarization energy can lead to a change of the polarization response by a factor of 2.

The flexoelectric effect, as a polarization response to a strain gradient, has been originally introduced as a purely bulk effect being viewed as a simple analogue of the piezoelectric effect \cite{Kogan1964,Bursian1974} (for a historical overview see e.g.\cite{Yudin2013}).
However, further  theoretical studies of the problem \cite{Tag1986} revealed the presence of unexpected contributions having no analogues in piezoelectricity. It was, for instance, realized that the flexoelectric response of a finite sample should be essentially influenced by effects controlled by the sample surface. This result, originally based on general arguments, recently received a direct confirmation based on first principles calculations \cite{Stengel2014}.

However, in the phenomenological framework one can single out the so-called \emph{static bulk flexoelectric effect}, which can be viewed as an analogue of the piezoelectric effect and others, additional contributions \cite{Yudin2013}.
It is this  effect that is customarily used for the description of the flexoelectricity-related phenomena in continuum theory of finite samples (see e.g. \cite{bib:Elis09,Sharma2008,Zhang2014}).
 It is usually addressed in terms of the following density of a thermodynamic potential \cite{bib:Elis09,Yudin2013}
\begin{equation}
\label{bulk-en}
{\cal F}_V = \frac{1}{2}g_{ijkl}P_{i,j}P_{k,l} +\frac{1}{2}a_{ij}P_iP_j  +
\frac{1}{2} f_{ijkl}(P_{k,l}u_{ij} - P_{k}u_{ij,l}) - E_iP_i \, ,
\end{equation}
where $E_i$, $P_i$, and $u_{ij}$ are  the Cartesian components of electric field,  polarization and strain tensor respectively, the suffixes  separated by a comma mean the corresponding spatial derivatives.
The flexoelectric effect is controlled here by a Lifshitz-type invariant containing the so-called flexocoupling tensor $f_{ijkl}$. Hereafter the Einstein summation convention is adopted.

For the description of the direct flexoelectric response of a finite sample, i.e. polarization response to a strain gradient, similar to the case of the dielectric response (c.f. \cite{kretschmer1979prb}), one should minimize with respect to the polarization the total thermodynamic potential of the sample, which  incorporates the surface polarization energy introduced by Kretschmer and Binder \cite{kretschmer1979prb}:
\begin{equation}
\label{en}
F = \int {\cal F}_V dV + \oint \frac{1}{2}a^S_{ij}P_iP_j dS \, .
\end{equation}
Here, in the last rhs term, the integration is done over the sample surface and the two-suffix quantity $a^S_{ij}$, controlling the surface polarization energy, is dependent on the orientation of the surface.

Minimization $F$ with respect to the polarization yields the equation of state
\begin{equation}
\label{bulk-eq}
a_{lk}P_l - g_{ijkl}P_{i,j,l} - f_{ijkl}u_{ij,l} - E_k = 0
\end{equation}
with the boundary conditions \cite{bib:Elis09,note2}
\begin{equation}
\label{surf-eq}
\left.
a^S_{lk}P_l + g_{ijkl}P_{i,j}n_l + \frac{1}{2}f_{ijkl}u_{ij}n_l
\right|_S = 0 \, ,
\end{equation}
where $n_i$ are the components of the unit vector  normal to the surface.

The framework presented above  enables the description of both dielectric and flexoelectric responses of a finite sample. For the geometry of the parallel plate capacitor the former was addressed in many papers (see e.g.\cite{kretschmer1979prb,Vendik2000,Tagantsev2006}).
A remarkable qualitative conclusion of this theoretical analysis is the difference between the dielectric response in the limiting cases where $a^S_{lk} =0 \,$ (the so-called free boundary condition) and $a^S_{lk} \to \infty$ (blocking boundary condition). 
In the "free" case, where the polarization surface energy vanishes, the effective value \cite{note1} of the  dielectric constant was found to be independent of the plate thickness $h$ and equal to its bulk value.
At the same time, in the "blocking" case the effective value of the  dielectric constant was found to be smaller than its bulk value, with the difference scaling as $1/h$.
An important feature of these results is that in the "thermodynamic limit" $h \to \infty$ the dielectric response becomes insensitive to the polarization surface energy.

In the present letter we  test the sensitivity of the flexoelectric response of a plate, conditioned by the static bulk contribution,  to the polarization surface energy.
To do this, we calculate it for the free and blocking cases to find a situation drastically different from that of the dielectric response. Specifically, we find that
(i) in the free  case  the response can be weaker that in the blocking one,
(ii) the difference between the cases does not disappear in the thermodynamic limit $h \to \infty$
and (iii) in the limit $h \to \infty$ the blocking case yields  the flexoelectric response twice that of the free case.

\begin{figure}[h]
\begin{center}
\includegraphics[width=8cm,keepaspectratio]{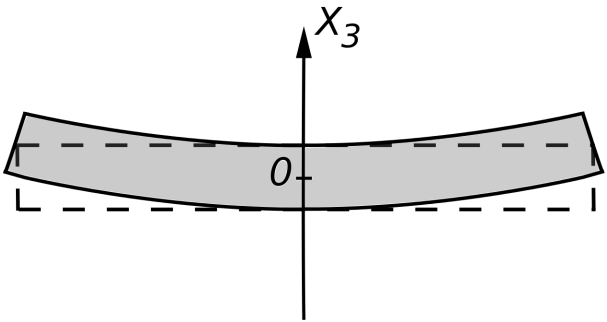}
\end{center}
\caption{A central cross-sections of plate: dashed line -- before bending and solid line  -- after.}
\label{fig1}
\end{figure}

Consider a thin (001) plate of a cento-symmetric cubic material of thickness $h$ normal the $X_3$ Cartesian coordinate, the origin being placed in its center  (Fig. \ref{fig1}). The plate is symmetrically bent,  so that the curvature of the plate $G$ is the same  in all cross-sections normal to it. Following   a thin plate approximation \cite{Landau} we present the stains  in the form
\begin{equation}
\label{strain}
u_{11}=u_{22}=X_3G;~~~ u_{33}=-2X_3\frac{c_{1122}}{c_{1111}}G;~~~ u_{12}=u_{23}=u_{13}=0,
\end{equation}
where $c_{1122}$ and $c_{1111}$ are the components of the tensor of elastic moduli. In the same approximation, in view of the geometry of the system and the symmetry of the material,
all variables depend only on the $X_3$ Cartesian coordinate and only  $P_3$ and $E_3$ components of the polarization and electric field, denoted hereafter as $P$ and $E$, respectively, are non-zero.

Using Eqs.~\eqref{bulk-eq} and ~\eqref{strain} and denoting the derivative with respect to $X_3$ as the prime, the polarization equation of state reads
\begin{equation}
\label{eq:difur}
P = P_\textrm{G}+\chi g P'' +\chi  E  \, ,
\end{equation}
where $g=g_{3333}$, $\chi=1/a_{33}$, and
\begin{equation}
\label{PG}
 P_\textrm{G} =  \chi\frac{c_{1111}f_{1122}-c_{1122}f_{1111}}{c_{1111}}2G.
\end{equation}
It is clear from Eq.~\eqref{eq:difur} that  $P_\textrm{G}$ signifies the polarization, induced by the plate bending,  if the polarization  were homogeneous and the electric field in it were zero.

For the general situation one should take into account the Poisson equation for the electrical displacement $D$, which we present in the form allowing  an adequate description of the depolarizing effects in ferroelectrics \cite{Tagantsev2006,Tagantsev2008}
\begin{equation}
\label{D}
 D= P +\varepsilon_\textrm{b} E ,
\end{equation}
where $\varepsilon_\textrm{b}$ is the background permittivity while $P$ is redefined as the "flexoelectric" contribution to the polarization.

Since all variables are functions of the coordinate  $X_3$ only, the Poisson equation implies $D={\rm const}$. So that
\begin{equation}
\label{Poisson}
 \langle P\rangle +  \varepsilon_\textrm{b}  \langle E\rangle=
 P + \varepsilon_\textrm{b} E ,
\end{equation}
where
\begin{equation}
\langle E\rangle = \frac{1}{h}\int\limits^{h/2}_{-h/2}EdX_3  \,~~~ \textrm{and} ~~~
\langle P\rangle = \frac{1}{h}\int\limits^{h/2}_{-h/2}PdX_3 .
\end{equation}
Combining Eqs.~\eqref{eq:difur} and ~\eqref{Poisson} we find
\begin{equation}
\label{master}
(1 +\chi/\varepsilon_\textrm{b}) P-\chi g P''=
 \chi/\varepsilon_\textrm{b}\langle P\rangle + P_\textrm{G} +\chi  \langle E\rangle.
\end{equation}
Equation \eqref{master} appended with the boundary conditions specify the problem.
Using Eq.~\eqref{surf-eq}, the latter read
\begin{equation}
\label{block}
\left. P \right|_{X_3=\pm h/2} = 0
\end{equation}
for the blocking case ($a^S_{lk} \to \infty$) and
\begin{equation}
\label{free}
 P'|_{X_3=  h/2} = - P'|_{X_3= - h/2}= - \frac{h}{4g\chi}P_\textrm{G}
\end{equation}
for the free case  ($a^S_{lk} =0 \,$).

The solutions to linear differential equation (\ref{master}), satisfying  boundary conditions (\ref{free}) or (\ref{block}), can be routinely   obtained.
However, being interested in $\langle P \rangle$ only and using the results available in the literature, one can evaluate the polarization response sought without  explicit solving of the differential equation, as presented below.

Consider the blocking case. The problem of the direct flexoelectric response of the shorted plate with the blocking boundary conditions, given by Eqs.~\eqref{master} and ~\eqref{block}, can be readily solved using results on the dielectric response of the plate with such boundary conditions. Indeed, as is clear from Eq.~\eqref{master} the flexoelectric response to the strain gradient, characterised  by a value of $P_\textrm{G}$, is equivalent to the dielectric response to the average field $\langle E\rangle=P_\textrm{G}/ \chi$, calculated neglecting ferroelectricity. Thus, the average polarization produced by the strain gradient can be cast in the form
\begin{equation}
\label{Pblock}
\langle P\rangle =P_\textrm{G}\,\frac{\chi_{\textrm{eff}}}{\chi} ,
\end{equation}
where $\chi_{\textrm{eff}}$ is the effective dielectric susceptibility of the system, i.e.  $\langle P\rangle/\langle E\rangle$ is calculated neglecting flexoelectricity.
Results of the earlier papers \cite{kretschmer1979prb,Vendik2000,Tagantsev2006} suggest
\begin{equation}
\label{chieff}
\chi_{\textrm{eff}} =\chi\,\frac{1-2\xi/h}{1+(2\xi/h)(\chi/\varepsilon_\textrm{b})} ,
\end{equation}
where
\begin{equation}
\label{xi}
\xi^2 =\frac{g\chi}{1+\chi/\varepsilon_\textrm{b}} .
\end{equation}
Equation (\ref{chieff}) is valid under the condition $h>>\xi$. Using typical values of $g$ in ferroelectrics, one finds that, depending on the value of $\chi/\varepsilon_\textrm{b}$, $\xi$ varies in the interval between the lattice constant and the correlation length of the material, which is about $\sqrt{g\chi}$ . Thus, this condition should hold for any realistic physical situation.
Equations ~\eqref{Pblock} and ~\eqref{chieff} implies
\begin{equation}
\label{blockf}
\langle P\rangle =P_\textrm{G}\,\frac{1-2\xi/h}{1+(2\xi/h)(\chi/\varepsilon_\textrm{b})}.
\end{equation}
Note that $\langle P\rangle =P_\textrm{G}$ in the "thermodynamic limit" $h \to \infty$.

In contrast to the blocking case, for the free case  the problem cannot be reduced to that for the dielectric response of a plate having vanishing surface polarization energy since the boundary conditions for the latter, $P'|_{X_3= \pm h/2} = 0$  are different from \eqref{free}.
However,     by averaging Eq.~\eqref{master}  over the plate thickness with $\langle E\rangle=0$,  we find a relationship
\begin{equation}
\label{average}
\langle P\rangle =P_\textrm{G} + \frac{g\chi}{h}
\left(
 P'|_{X_3=  h/2} - P'|_{X_3= - h/2}
\right) ,
\end{equation}
which being combined with Eq.~\eqref{free} yields the flexoelectric response of the shorted plate as
\begin{equation}
\label{freef}
\langle P\rangle =P_\textrm{G}/2.
\end{equation}
It is instructive to recall that in the free case $\chi_{\textrm{eff}} =\chi$ \cite{kretschmer1979prb,Tagantsev2006}.

The above results reveal  a drastic difference between the impact of the surface polarization energy on the dielectric and flexoelectric responses. The most striking is the fact that in the thermodynamic limit $h \to \infty$ the flexoelectric response is two times different between the blocking and free cases while there is no such difference for the dielectric response.  
The behaviour of the dielectric response is readily expected: the impact of a surface perturbation (polarization surface energy) vanishes in the thermodynamic limit. 
Meanwhile, even in the thermodynamic limit the contribution of the \textit{bulk flexoelectric effect} to the direct flexoelectric response was found to be sensitive to the surface perturbation. 
Actually, a similar phenomenon was identified in the case of the contribution of the  so-called surface piezoelectricity to the  total flexoelectric response of the sample \cite{Yudin2013,bib:TagYu12}.
\begin{figure}[t]
\begin{center}
\includegraphics[width=8 cm,keepaspectratio]{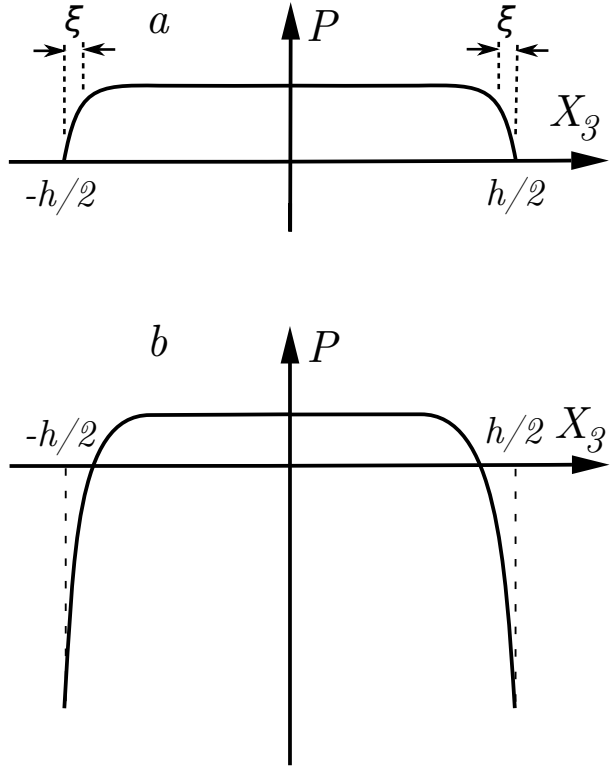}
\end{center}
\caption{Schematics for the distribution of polarization across plate thickness, caused by its bending: (a)  blocking boundary conditions, (b) free boundary conditions.
$h$ and $\xi$ and the plate thickness and the spatial scale given by Eq.~\eqref{xi}, respectively.  }
\label{fig2}
\end{figure}
Like in the case of surface piezoelectricity, under short-circuited  condition, the strain gradient induces  surface dipoles which in turn create an electric field in the body of the plate.
The formation of such dipoles is seen in Fig.~\ref{fig2}.
The surface dipoles form in  both blocking and free cases.
However, in the  former, the surface dipole moment is independent of the plate thickness $h$  while in the latter it scales as $h$.
In the free case, this gives rise to a non-vanishing impact of the surface perturbation in the thermodynamic limit.

It is worth to mentioning that, in view of basic thermodynamic arguments, one expects that the  two-times difference between the direct flexoelectric effects in the blocking and free cases must imply the same difference between the converse flexoelectric effects (bending of the plate in a homogeneous electric field) in these cases. This expectation is in full agreement with the calculations of the field-induced plate bending (converse flexoelectric effect) controlled by the static bulk flexoelectric effect performed for the blocking \cite{bib:TagYu12} and free \cite{bib:YurkovPSS14} cases. Indeed, the converse effect was found to be twice  stronger in the blocking case.

 One should also comment  on the applicability of the continuum approach to the problem addressed.  The typical spatial scale for the polarization variation is $\xi$ as clear from  Eqs.~\eqref{master}, \eqref{xi}, and Fig.~\ref{fig2}.  The applicability of the continuum approach  requires that $\xi$ should appreciably exceed the lattice constant of the material. This condition is readily met for the so-called weak ferroelectrics \cite{Tagantsev1987,Tagantsev2008}, where $\varepsilon_\textrm{b}$ is of the order of $\chi$ and $\xi$ is about the correlation length that, in turn, by definition appreciably exceeds the lattice constant. In the case of "regular" proper ferroelectrics, where $\varepsilon_\textrm{b}<<\chi$, $\xi$ is expected be about the lattice constant of the material.  This pushes the continuum approach to the edge of its applicability range.
 However, one may expect that the qualitative conclusions of the presented analysis will hold in this case also.

 In conclusion, we have shown that the direct flexoelectric response of the plate associated with the \emph{static bulk flexoelectric effect} is very sensitive to the surface polarization energy.
 In contrast to the polarization response, a modification of the surface polarization energy can result  a variation of the flexoelectric response by a factor of 2.
 Remarkably, the impact of such surface perturbation holds in the thermodynamic limit.
 The reported results suggest that any modelling of electromechanical finite systems involving flexoelectricity should take into account the surface polarization energy and the modification of the polarization boundary conditions associated with the flexoelectric coupling.  Nowadays such modellings customarily neglect these issues.

\begin{acknowledgments}
This project was supported by the Swiss National Science Foundation.
The Research was also supported  by grant of the government of the Russian Federation 2012-220-03-434.
\end{acknowledgments}


\begin{thebibliography}{99}

\bibitem{kaganov1972}
M. I. Kaganov and A. N. Omelyanc, Sov. Phys. JETP \textbf{34}, 895 (1972).

\bibitem{kretschmer1979prb}
R. Kretschmer and K. Binder, Phys. Rev. B, \textbf{20}, 1065 (1979).

\bibitem{Junquera2003}
J. Junquera and P. Ghosez, Nature (London) \textbf{422}, 506 (2003).

\bibitem{Stengel2009}
M. Stengel, D. Vanderbilt, and N. A. Spaldin, Nature Mater. \textbf{8}, 392 (2009).

\bibitem{note1}
It is calculated from the derivative of the charge of the plates of a parallel plate capacitor containing the film with respect to the applied voltage.

\bibitem{Tagantsev2006}
A.K.~Tagantsev, G.Gerra,   J. Appl. Phys. \textbf{100}, 051607 (2006).

\bibitem{Kogan1964}
S. M. Kogan, Sov. Phys. Solid State \textbf{5}, 2069 (1964).

\bibitem{Bursian1974}
E. V. Bursian and N. N. Trunov, Sov. Phys. Solid State \textbf{16}, 760 (1974).

\bibitem{Yudin2013}
P. V. Yudin and A. K. Tagantsev, Nanotechnology \textbf{24}, 432001 (2013).

\bibitem{Tag1986}
A.K. Tagantsev, Physical Review B, \textbf{34}, 5883 (1986).

\bibitem{Stengel2014}
M. Stengel,  Physical Review B, \textbf{90}, 201112, (2014).

\bibitem{bib:Elis09}
E.A.~Eliseev, A.N.~Morozovska, M.D.~Glinchuk, R.~Blinc, Phys.~Rev.~B {\bf 79}, 165433 (2009).


\bibitem{Sharma2008}
M. S. Majdoub, P. Sharma, T. Cagin, Phys.~Rev.~B {\bf 77}, 125424 (2008).

\bibitem{Zhang2014}
Z. Zhang, Z. Yan, and L. Jiang, J.~Appl.~Phys. {\bf 116}, 014307 (2014)

\bibitem{note2}
These boundary conditions were originally derived in \cite{bib:Elis09}, where these are given with a misprint in the order of suffices of the  tensor $g_{ijkl}$, corrected in the present paper.

\bibitem{Vendik2000}
O.G.~Vendik, S.P.~Zubko, J.~Appl.~Phys. {\bf 88}, 5343 (2000).

\bibitem{Landau}
L. D. Landau and E. M. Lifshitz, Theory of Elasticity (Pergamon Press, Oxford, 1975).



\bibitem{Tagantsev2008}
A. K. Tagantsev, Ferroelectrics, {\bf 375},  19 (2008).

\bibitem{bib:TagYu12}
A.K.~Tagantsev, A.S.~Yurkov, J.~Appl.~Phys. {\bf 112}, 044103 (2012).


\bibitem{bib:YurkovPSS14}
A.S.~Yurkov, Phys.~Solid~State  {\bf 57}, 460 (2015), (in press). 

\bibitem{Tagantsev1987}
A. K. Tagantsev, I. G. Sinii, and S. D. Prokhorova, Izvestiya Akademii Nauk SSSR Seriya Fizicheskaya
{\bf 51}, 2082 (1987).


\end{thebibliography}
\end{document}